# Speckle Optical Tweezers: Micromanipulation with Random Light Fields

Giorgio Volpe,[*a] Lisa Kurz,[a] Agnese Callegari,[b] Giovanni Volpe[b,c] and Sylvain Gigan[a]

a. Institut Langevin, UMR7587 of CNRS and ESPCI ParisTech, 1 rue Jussieu, 75005 Paris, France

b. Soft Matter Lab, Physics Department, Bilkent University, Cankaya, 06800 Ankara, Turkey

c. UNAM – National Nanotechnology Research Center, Bilkent University, Cankaya, 06800 Ankara, Turkey

*Corresponding author: giorgio.volpe@espci.fr

## Abstract

Current optical manipulation techniques rely on carefully engineered setups and samples. Although similar conditions are routinely met in research laboratories, it is still a challenge to manipulate microparticles when the environment is not well controlled and known a priori, since optical imperfections and scattering limit the applicability of this technique to real-life situations, such as in biomedical or microfluidic applications. Nonetheless, scattering of coherent light by disordered structures gives rise to speckles, random diffraction patterns with well-defined statistical properties. Here, we experimentally demonstrate how speckle fields can become a versatile tool to efficiently perform fundamental optical manipulation tasks such as trapping, guiding and sorting. We anticipate that the simplicity of these "speckle optical tweezers" will greatly broaden the perspectives of optical manipulation for real-life applications.

# Introduction

Since their introduction in the 1970s,[1,2] optical tweezers have been widely applied to non-invasively manipulate micro- and nano-objects, such as cells, organelles and macromolecules.[3-5] They have, therefore, gained increasing importance as tools in microbiology and biophysics both for fundamental studies[6] and for more advanced applications such as optical sorting and optical delivery.[3,7-8] In particular, the development of techniques based on reconfigurable spatially extended patterns of light, such as multiple traps[3,9-12] or periodic potentials,[13-17] offers the promise of high throughput optical methods to be applied both in static and moving fluids. Also, particles' delivery, trapping and manipulation over extended areas was demonstrated near a surface employing the evanescent fields associated, for example, to surface plasmons[18] or to optical waveguides.[19]

Most of current optical manipulation techniques rely either on carefully engineered optical systems or advanced fabrication tools. Although such conditions can be routinely met in research laboratories, similar requirements, sometimes very stringent, limit the applicability of these techniques, e.g., to biomedical and microfluidic applications, where simplicity, low-cost and high-throughput are paramount. Moreover, one major challenge common to all these techniques is the light scattering occurring in optically complex media, such as biological tissues, turbid liquids and rough surfaces, which naturally gives rise to apparently random light fields known as *speckles*.[20] Earlier experimental works showed trapping of atoms and particles in a gas by high-intensity speckle light fields,[21-24] while both static and time-varying speckle fields were related to the emergence of anomalous diffusion in colloids.[25-29] Recently, we derived a theory to describe the motion of a Brownian particle in a speckle light field which allowed us to demonstrate numerically how a speckle field can be used to control the motion a Brownian particle in the limit of particles much smaller than the light wavelength (dipole approximation).[29] However, apart from these previous studies, the intrinsic randomness of speckle patterns is largely considered a nuisance to be minimized for most purposes in optical manipulation.[30-31]

Here, we experimentally demonstrate a novel technique for the collective manipulation of micrometer-sized particles in microfluidic flows based on extended static and time-varying speckle light fields. Just relying on the statistical interaction between the particles and the underlying optical potential, these *speckle optical tweezers* allow us to perform important optical manipulation tasks such as sieving, guiding and sorting within a microfluidic channel.

## Methods

The speckle optical tweezers setup is schematically depicted in Fig. 1a. Aqueous dispersions of colloidal spheres are driven by a syringe pump with adjustable infusion flow rate (Harvard Apparatus Pump 11 Elite) through a microfluidic channel. The speckle light pattern for their optical manipulation (Fig. 1b) is generated by coupling a laser beam (Coherent Verdi, maximum power 5W, $\lambda = 532$ nm) into a multimode optical fiber (core diameter 105 μm, NA = 0.22). The random appearance of speckle light patterns is the result of the interference of a large number of optical waves with random phases, corresponding to different eigenmodes of the fiber. More generally, speckle patterns can be generated by different processes: scattering of a laser on a rough surface, multiple scattering in an optically complex medium, or, like in this work, mode-mixing in a multimode fiber.[32] The method chosen in this work provides some practical advantages over other methods, namely the generation of homogeneous speckle fields over controllable areas, flexibility and portability in the implementation of the device, as well as higher transmission efficiency. In our setup, the fiber output is brought in close proximity of the upper wall of the microfluidic channel by a micrometric two-axis mechanical stage that also guarantees the possibility of translating the speckle in the direction perpendicular to the fluid flow. Optical scattering forces push the particles in the direction of light propagation towards the lower wall of the microfluidic channel, so that they effectively confine the particles in a quasi two-dimensional space.[1] The particles are then tracked by digital video microscopy[33] on a color CMOS camera. The incoherent illumination for the tracking is provided by a LED at $\lambda = 625$ nm coupled into the same fiber using a dichroic mirror. Fig. 1c shows the normalized spatial autocorrelation function of a typical speckle pattern interacting with the particles (Fig. 1b), whose full width half maximum (FWHM = $2.10 \pm 0.24$ µm) provides an estimation of the average speckle grain size, as defined by the diffraction process that generates the speckle pattern itself.[20,34]

## Results and Discussion

We start by considering the simplest case, e.g., the motion of an isolated silica bead (diameter D = $2.06 \pm 0.05$ µm, refractive index $n_p = 1.42$) in a static speckle pattern and without fluid flow. As shown by the trajectory (solid line) in Fig. 2a, when the average speckle intensity is relatively low ($\langle I \rangle = 0.12$ µW/µm²), the particle is virtually freely diffusing. As the intensity increases (Fig. 1b, $\langle I \rangle = 1.43$ µW/µm²), the particle gets metastably trapped in the speckle grains, while it can still jump from one grain to the next from time to time.[35] Finally, for even higher intensities (Fig. 1c, $\langle I \rangle = 5.77$ µW/µm$^2$), the particle remains trapped in one of the speckle grains.

To gain further insight on the underlying physics, we calculated the force field acting on a silica particle moving in a simulated speckle pattern (Fig. 2d). For the calculation of the optical forces, as the particle size is significantly larger than the light wavelength, we used a ray optics approach[36] and, for the simulation of the particle motion, we employed Brownian dynamics simulation.[37] The details of the simulated trajectories (Figs. 2e-g) of silica particles moving in speckle fields of the same average intensity as in Figs. 2a-c show very good agreement with the experimental data. In general, the motion of a Brownian particle in a static speckle field is the result of random thermal forces and deterministic optical forces.[29] Optical gradient forces are the dominant deterministic forces acting on dielectric particles whose size is comparable or smaller than the average speckle grain, and they attract particles with high-refractive index towards the intensity maxima of the optical field.[29,38] As a particle moves in the speckle field, the optical force acting on it changes both in magnitude and direction with a characteristic time scale that in first approximation is inversely proportional to the average speckle intensity.[29]

Since the optical forces exerted on a particle depend on the particle's physical parameters, e.g., size, refractive index and shape,[38] a static speckle pattern can be employed to realize a *speckle sieve* in the presence of flow (Fig. 3). In Figs. 3a-f, as an aqueous dispersion containing two kinds of particles of similar diameter, i.e., D ≈ 2 µm, but different refractive index (silica, $D = 2.06 \pm 0.05$ µm and $n_p = 1.42$, and melamine, $D = 2.05 \pm 0.04$ µm and $n_p = 1.68$) flows from left to right at $V_f = 3.01 \pm 0.12$ µm/s, a static speckle pattern efficiently holds back the particles with higher refractive index (melamine) while the ones with lower refractive index (silica) go through almost unaffected (Figs. 3a-f). These qualitative considerations can be made more precise by calculating the average particle speed $\langle V_p \rangle$ in the microfluidic speckle sieve. As shown in Fig. 3g, when the laser is off, the particles are flowing at the speed of the surrounding medium because of the fluid laminar flow.[39] As the speckle intensity increases, $\langle V_p \rangle$ converges to zero: for a given class of particles, this convergence happens for higher speckle intensities when the fluid flows faster; accordingly, for a given fluid flow, the higher the particle refractive index is, the lower is the requirement on the speckle intensity, thus allowing one to sieve particles with different physical characteristics, as in Figs. (a-f). Fig. 3h shows that similar conclusions hold when the selection parameter is the particle size rather than its refractive index. Interestingly, the physical characteristics of the particle that are held back can be dynamically adjusted by changing the intensity of the speckle pattern.

Time-varying speckle patterns are also very versatile tools to control the motion of Brownian particles, thus setting the stage to perform optical manipulation tasks such as guiding particles in a particular direction, despite the randomness of the illumination.[29] In Figs. 4a-c, the speckle pattern shifts first slowly in the direction indicated by the white arrow, which exerts a strong adiabatic drag on a melamine particle, and then fast back to the initial position with little effect on the position of the particle since the movement is too fast for the particle to follow. Due to the much lower optical forces, the position of a nearby silica particle with similar size remains almost unchanged during the whole time. Repeating this cycle as shown in Fig. 4d is sufficient to realize a Brownian ratchet (Fig. 4e):[40] in 26 s the melamine particle is dragged by $\approx 35\mu m$ in the direction of the speckle pattern shift, while the particle's trajectory in the perpendicular direction remains almost unaffected; similarly, the silica particle is also dragged in the same direction, albeit much less efficiently ($\approx 11\mu m$). In this experiment, the shift of the speckle pattern is induced by moving the fiber with a mechanical translation stage. Interestingly, a small speckle pattern translation up to a few micrometers can also be implemented capitalizing on the speckle property known as memory effect:[41-42] for a speckle pattern generated by a thin sample, a small tilt of the illumination, easily achievable, e.g., with a galvanometric mirror or an acousto-optic deflector, entails a small spatial translation of the speckle pattern. The speckle ratchet that we propose here, therefore, can also be implemented in real situations thanks to the speckle memory effect.

In the presence of a flow, we can capitalize on the guiding capability of speckle patterns in order to perform optical sorting and fractionation.[7,11-12] In a configuration similar to the one for the speckle sieve, a shifting speckle can be used to realize a *speckle sorter*, where a force perpendicular to the flow is selectively exerted on different classes of particles, so that each kind is deflected at a different angle (Fig. 5).

As shown in Fig. 5a, when the laser is off, the particles (both melamine and silica of the same size) are flowing in the direction of the surrounding medium because of the fluid laminar flow.[39] As the average speckle intensity increases (Figs. 5b-c), the particles start being deflected from the direction of the fluid flow: for a given class of particles, the deflection angle grows with the speckle intensity, while, for a given intensity, it grows with the refractive index of the particle. This qualitative behavior is independent of the flow speed, although higher average intensities are needed to achieve comparable deflection angles at higher speeds (Figs. 5d-g). Figs. 5h-n show that similar conclusions hold when the selection parameter is the particle size rather than its refractive index. Interestingly, the resolution of this optical fractionation is only limited by the size of the speckle field, i.e., the longer the speckle field the higher the sensitivity in particle's size or refractive index.[29]

## Conclusions

In conclusion, we have experimentally demonstrated a novel technique for the optical manipulation of microparticles in microfluidic flows based on static, time- or space-varying speckle fields. The required optical intensities are comparable to those reported in similar studies where the force field was generated either with holographic optical traps or with periodic potentials.[3,7-17] Moreover, an additional advantage of speckle patterns is that they are also intrinsically wide-field so that they have the potential of sorting many particles in parallel in a broader microfluidic chamber, where flow speed is strongly reduced. Our technique, beyond demonstrating that random potentials are a valid alternative to more regular potentials for the purpose of optical manipulation, offers some additional advantages to current optical manipulation techniques,[3,7-17] such as intrinsic robustness to noise and aberrations from the optics and the environment. Finally, the use of random optical potentials over periodic ones has the advantage of requiring very simple optical setups as well as a very low degree of control over the experimental environment, thus being readily compatible with optical delivery, lab-on-a-chip or in-vivo applications inside scattering tissues, where light propagation naturally leads to the formation of speckle patterns, without recurring to wavefront shaping.[30-31]

# Acknowledgements

The authors acknowledge Fabien Bertillot for his help in the development of the initial setup, and thank Marie Leman and Andrew Griffiths for discussion. Giovanni Volpe was partially supported by Marie Curie Career Integration Grant (MC-CIG) PCIG11 GA-2012-321726. Sylvain Gigan acknowledges funding from Agence Nationale de La Recherche (ANR-JCJC-ROCOCO), the City of Paris (Programme Emergence) and the European Research Council (under grant N°278025).

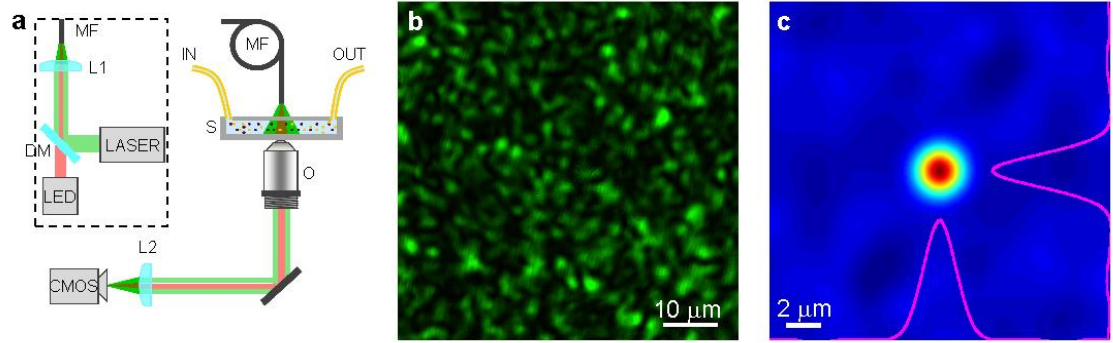

**Figure 1: Speckle optical tweezers setup.** (a) Schematic of the speckle optical tweezers setup. A laser beam (λ = 532 nm) and incoherent light from an LED (λ = 625 nm) are coupled into a multimode optical fiber (105-μm core, NA = 0.22) making use of a dichroic mirror (DM) and a lens (L1). The fiber delivers the light to a microfluidic channel (S) where aqueous dispersions of particles are flown by a syringe pusher. The fiber output is mounted on a two-axis mechanical stage, which guarantees the possibility of translating the speckle vertically and perpendicularly to the flow. The particles' trajectories are tracked by digital video microscopy using the image projected by a microscope objective (20X, NA = 0.5) and a tube lens (L2) onto a color CMOS camera. (b) A typical speckle pattern for optical manipulation as observed on the camera and (c) its normalized spatial autocorrelation function, which permits us to characterize the average speckle grain size as the FHWM of the autocorrelation along the axes (solid lines).

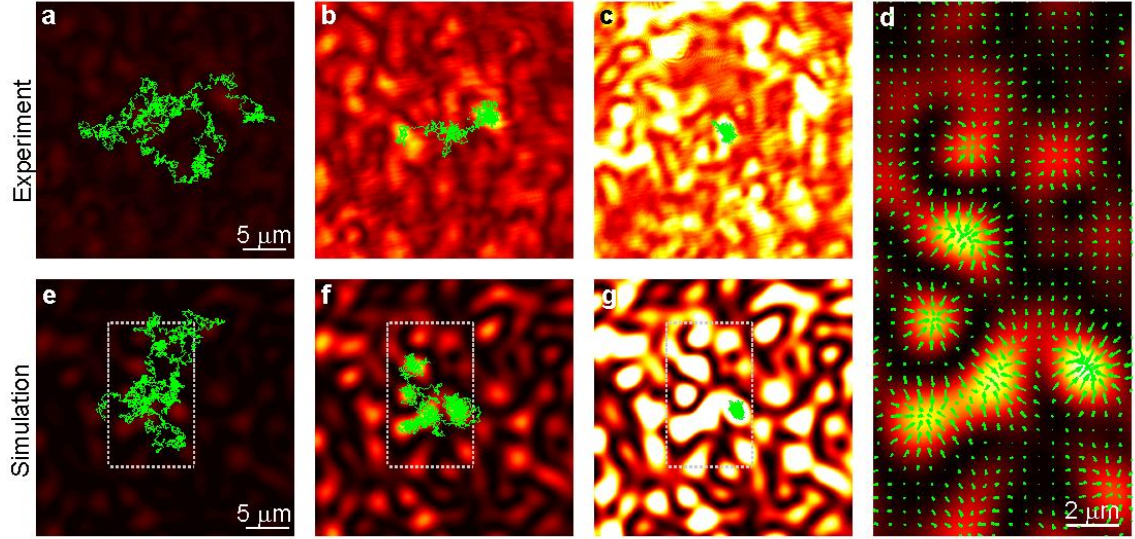

**Figure 2: Optical forces in a static speckle field.** (a-c) The experimental trajectories (solid lines) show progressive confinement of a silica bead (D = 2.06 ± 0.05 μm, $n_p$ = 1.42) in water ($n_m$ = 1.33) as a function of the increasing average speckle intensity, respectively ⟨I⟩ = 0.12 μW/μm² in (a), ⟨I⟩ = 1.43 μW/μm² in (b), and ⟨I⟩ = 5.77 μW/μm² in (c). The backgrounds are the corresponding images of the speckle patterns generated by mode-mixing in a multimode optical fiber. (d) Calculated force field (arrows) exerted on a silica bead in a simulated speckle pattern (background). (e-g) Corresponding simulated trajectories (solid lines) of silica particles moving in speckle fields of the same average intensity as in (a-c). The dashed lines delimit the area corresponding to the force field distribution in (d). The average calculated force exerted by the speckle field is (e) ⟨F⟩ = 0.14 fN, (f) ⟨F⟩ = 1.82 fN, and (g) ⟨F⟩ = 7.3 fN. All trajectories are recorded or simulated during 420 s.
13

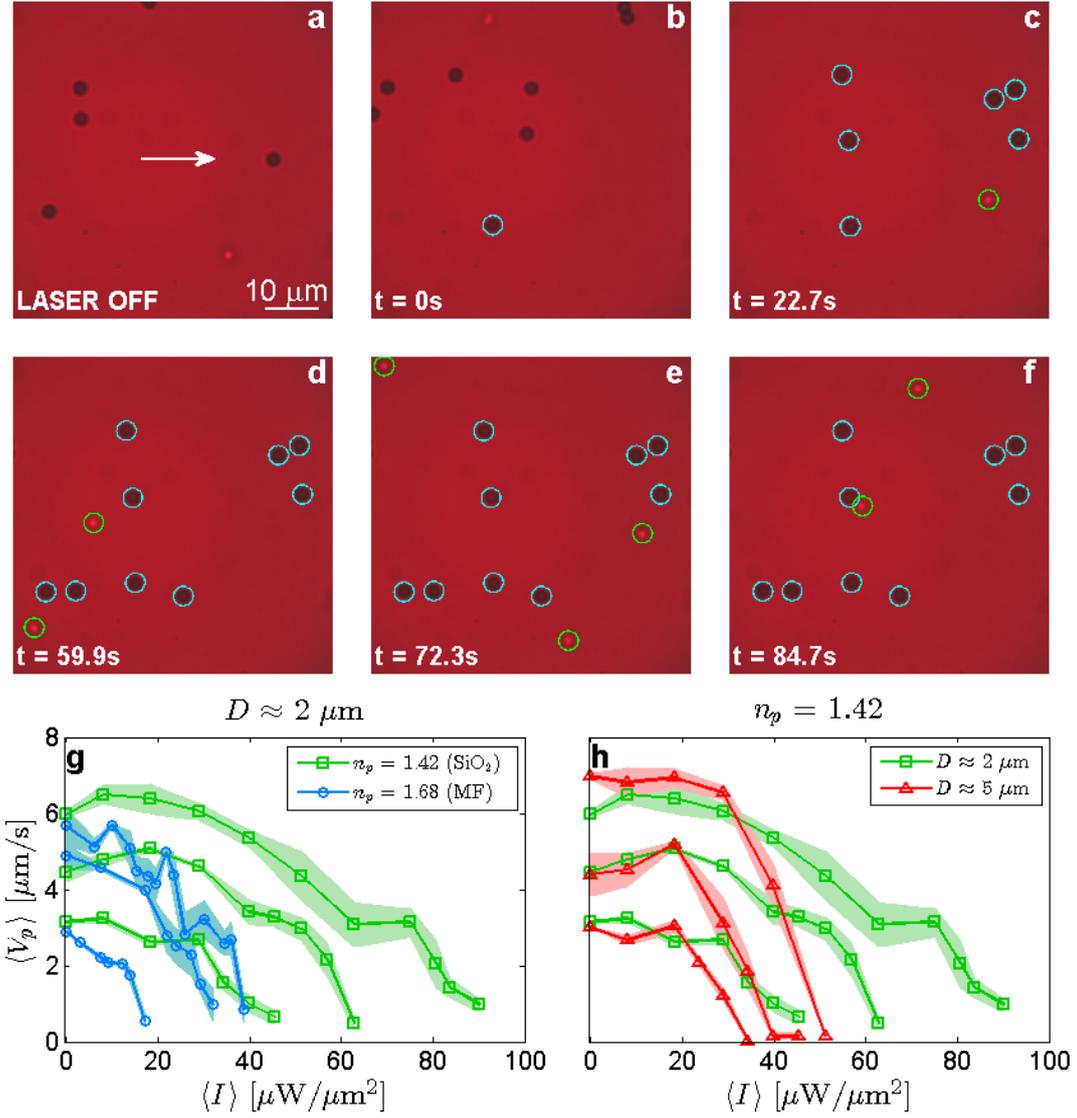

**Figure 3: Sieving in a microfluidic flow by a static speckle field.** (a-f) Time-lapse snapshots of the flow of two classes of particles with similar diameter D ≈ 2 μm but different refractive index in a microfluidic speckle sieve (flow speed $V_f$ = 3.00 ± 0.12 μm/s): silica (brighter particles, D = 2.06 ± 0.05 μm and $n_p$ = 1.42) and melamine (darker particles, D = 2.05 ± 0.04 μm and $n_p$ = 1.68). The arrow in (a) indicates the direction of the flow. A static speckle pattern (on from (b), $\langle I \rangle$ = 21.9 μW/μm$^2$), traps the particles with higher refractive index (blue circles) while it lets the particles with lower refractive index (green circles) go away with the flow. (g-h) Comparison of the average particle speed $V_p$ in the speckle sieve (g) for particles of similar diameter (D ≈ 2 μm) but different refractive index (green squares, $n_p$ = 1.42, and blue circles, $n_p$ = 1.68), and (h) for particles of similar refractive index ($n_p$ = 1.42), but different diameter (green squares, D = 2.06 ± 0.05 μm, and red triangles, D = 4.99 ± 0.22 μm ), as a function of the average speckle intensity and of the fluid flow ($V_f$ = 3.01 ± 0.12 μm/s, $V_f$ = 4.58 ± 0.26 μm/s and $V_f$ = 6.20 ± 0.68 μm/s). The shaded areas represent one standard deviation around the average values.

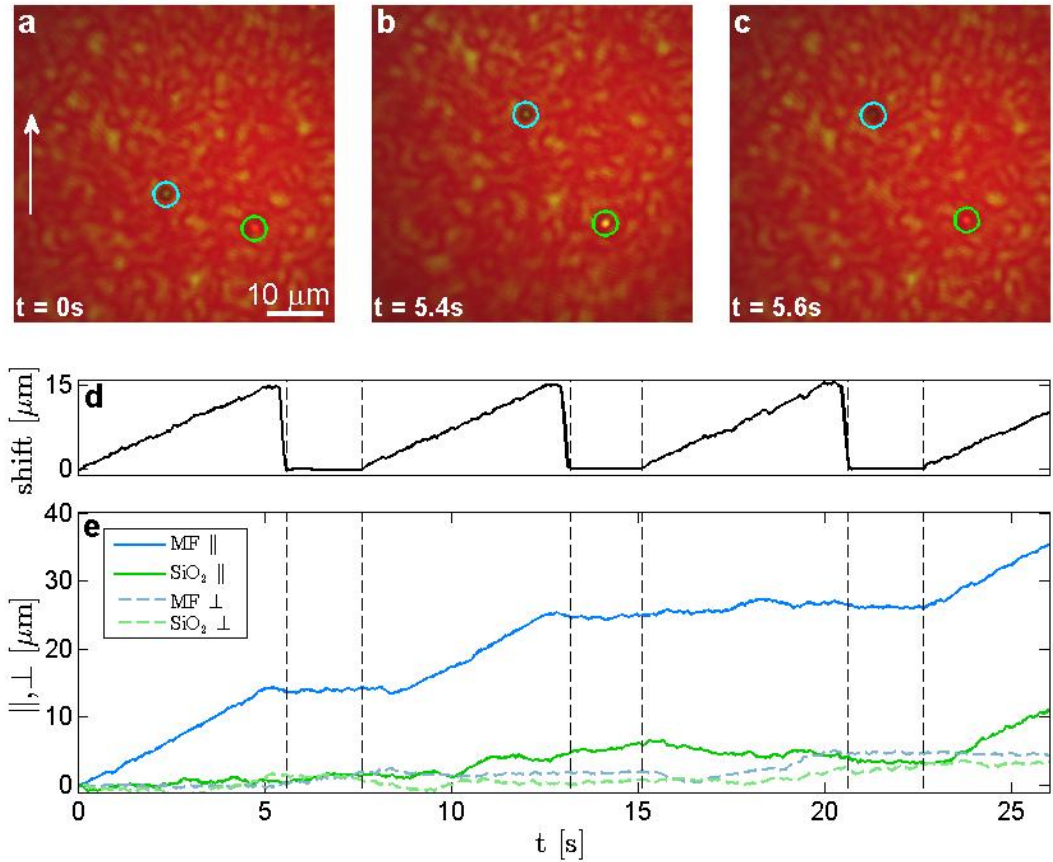

**Figure 4: Guiding by a ratcheting speckle.** (a-c) Time-lapse snapshots of the motion of a melamine particle (blue circles) and a silica particle (green circles) with similar diameter D ≈ 2 μm in a ratcheting speckle in the absence of flow ($\langle I \rangle$ = 7.85 μW/μm$^2$). The shift of the speckle, which is visible in the background, is induced by dragging the fiber with a mechanical stage first (from (a) to (b)) slowly in the direction of the arrow shown in (a) and then (from (b) to (c)) fast back. (d) Speckle pattern shift as tracked on a speckle grain and (e) particle displacements as a function of time in the direction parallel to the speckle pattern shift (solid blue and green lines) and in the orthogonal direction (dashed blue and green lines), respectively for the melamine and the silica particle. The speckle pattern repeatedly shifts first slowly in the positive

direction and then fast to the initial position in 5.6 s cycles. The dashed lines delimit the time of absence of motion due to the motor backlash.

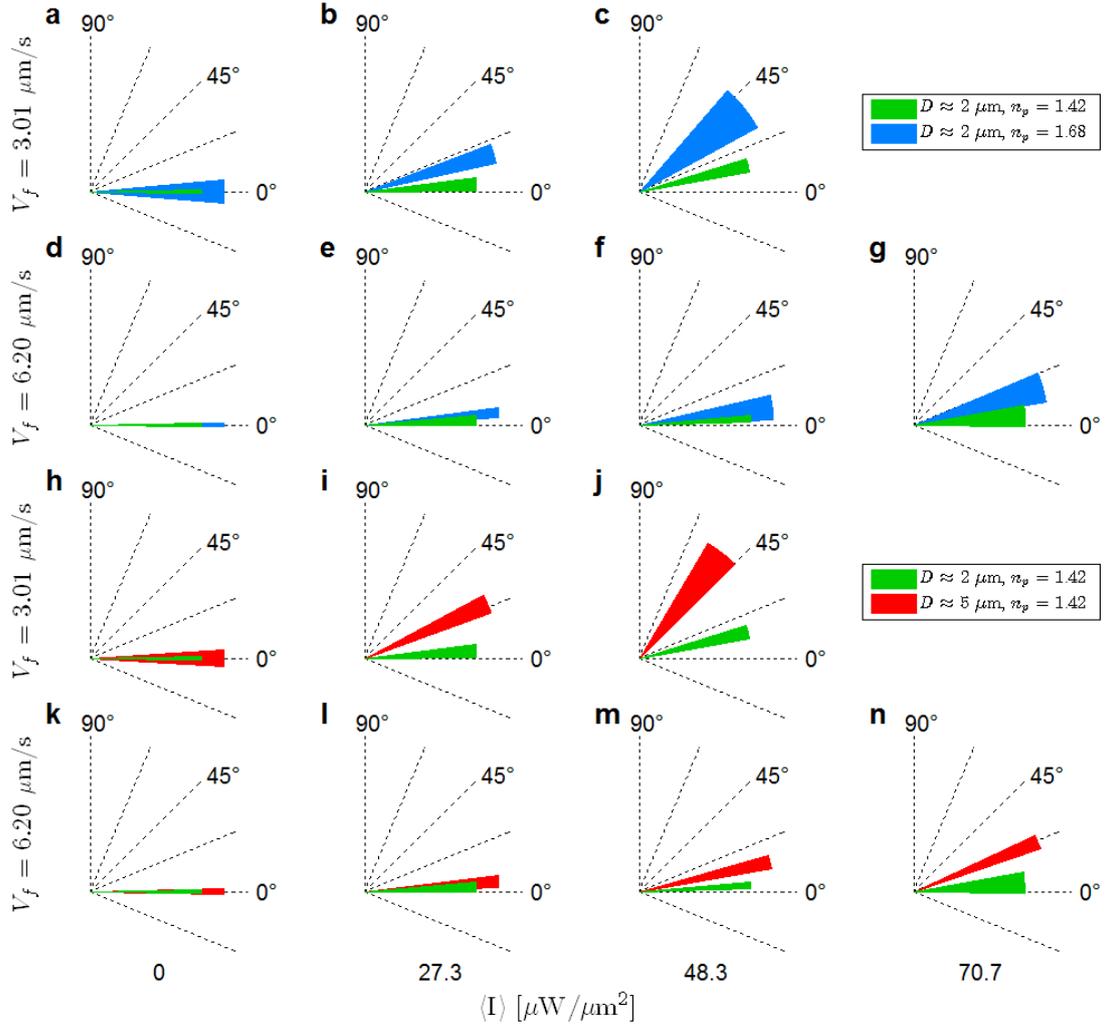

**Figure 5: Sorting in a microfluidic flow by a ratcheting speckle field.** (a-g) Angular distribution of two classes of particles with similar diameter (D ≈ 2 μm) but different refractive index ($n_p$ = 1.42, green areas, and $n_p$ = 1.68, blue areas) in a microfluidic speckle sorter for increasing flow speeds $V_f$ ($V_f$ = 3.01 ± 0.12 μm/s from (a) to (c) and $V_f$ = 6.20 ± 0.68 μm/s from (d) to (g)) and average speckle intensities ⟨I⟩. The flow is directed along the 0° line, while the speckle shift is directed along the 90° line. The areas represent one standard deviation of the particle spread around the average value. (h-n) Same as (a-g) using as selection parameter the particle size (D = 2.06 ± 0.05 μm, green areas, and D = 4.99 ± 022 μm, red areas) rather than their refractive index, here kept constant ($n_p$ = 1.42).